\documentclass[a4paper]{book}

\usepackage{style/infrastructure}
\usepackage{style/layout}
\usepackage{style/generic}
\ShowVersions

\begin{document}
   \SetAuthor{H.A. (Erik) Proper}
   \SetTitle{Work Systems Modeling Library}
   \pagestyle{empty}

\begin{titlepage}
   \begin{center}
      \mbox{~~}

      \vspace{2cm}

      {\Huge \Title}

      \vspace{1cm}
      
      {\LARGE \Author}

      {\large \TheVersion}

      \vspace{2cm} 

      \centerline{\ImageBox{8cm}{vitruvius}}

      \vspace{1cm}

      {\large \DaVinciSub}

      \vspace{1cm}
   \end{center}
\end{titlepage}

\begin{titlepage}
   \setcounter{page}{3}
   \begin{center}
      \mbox{~~}

      \vspace{2cm}

      {\Huge \Title}

      \vspace{1cm}

      {\LARGE \Author}
   \end{center}
\end{titlepage}

\mbox{~~}
\vfill
\Colophon
\newpage

\pagestyle{headings}
\tableofcontents

      \chapter{Introduction}
\Version{06-09-05}
As the title suggests, we are focussing on a library for the modeling of 
\PDN{work-system}. In line with~\cite{1999-Alter-WorkSystems,
2002-Alter-WorkSystems} we regard a \DN{work-system} as:
\begin{Definitions}
   An \DN{open-active-system} in which \PDN{actor} perform processes using
\DN{information}, technologies, and other resources to produce products and/or
services for internal or external actors.

\end{Definitions}
\UPDN{information-system} as well as \PDN{organization} are regarded as
specialisations of \PDN{work-system}. More precisely we define
\begin{Definitions}
   A special kind of \DN{system}, being normally active and open, and comprising
the conception of how an \DN{organization} is composed and how it operates
(i.e.\ performing specific actions in pursuit of organizational goals, guided
by organizational rules and informed by internal and external communication),
where its \DN{systemic-property} are that it responds to (certain kinds of)
changes caused by the system \DN{environment} and, itself, causes (certain
kinds of) changes in the system environment.

   A \DN{sub-system} of an \DN{organizational-system}, comprising the conception
of how the communication and information-oriented aspects of an
\DN{organization} are composed and how these operate, thus leading to a
description of the (explicit and/or implicit) commu\-nication-oriented and
information-providing actions and arrangements existing within the
\DN{organizational-system}.

   A \DN{sub-system} of an \DN{information-system}, whereby all activities within
that sub-system are performed by one or several computer(s).	

\end{Definitions}
Modeling of \PDN{work-system} occurs for all sorts of reasons. Requirements
need to be expressed. A pre-existing situation may need to be charted and
analyzed. Early design decisions may be captured using
\PDN{architecture-principle}.  Detailed design may be worked out. We all regard
these activities as essentially being forms of \DN{modeling}. In the
\PDN{work-system} \DN{modeling} library, we consider \DN{work-system}
engineering from a \DN{modeling} perspective. 

In the field of \DN{work-system} engineering, a whole plethora of modeling
\PDN{method} is available to system engineers and architects. Each of these
\PDN{method} can be used to \DN{model} some (aspects) of a domain related to an
existing and/or a planned \DN{work-system}.  The aspects may refer to
requirements, architecture, design, processing, data, etc, etc. In other words,
these methodes are essentially all intended to model different aspects of work
systems and/or their context.

The aim of the \PDN{work-system} \DN{modeling} library (\DN{WSML}) is to bring
together methodical knowledge concerning the \DN{modeling} of \PDN{work-system}.
This knowledge may, for example, pertain to:
\begin{itemize}
  \item the syntax and/or semantics of \DN{modeling} \PDN{technique} to be used, 
  \item guidelines or procedures describing how to produce the different models, 
  \item frameworks defining different foci for viewing a \DN{work-system}, 
  \item partial models (acting as pre-fabricated building blocks for more
  complex models) such as patterns and reference models, 
  \item formal associations between models such as transformations (for example
  mapping from a conceptual database schemas to a relational database schemas),
  refinements (from high level design to detailed design), etc.
\end{itemize}
To refer to any of such `fragments' of methodical knowledge, we shall use the
term \DN{method-fragment}, which we define as:
\begin{Definitions}
   An identifiable part of methodical \DefinitionName{knowledge}{ls}.

\end{Definitions}
The \DN{WSML} will be filled with \PDN{method-fragment} pertaining to
\DN{modeling} of \PDN{work-system}.

The motivations behind the \DN{WSML} are threefold:
\begin{itemize}
   \item Our research group is involved in research into \DN{work-system}
   \DN{modeling} in the broadest sense.  Having a library of
   \PDN{method-fragment} pertaining to the \DN{modeling} of \PDN{work-system},
   will aid us in our research activities. Even more, the high level structures
   for structuring the library, as well as the classification scheme to
   classify the available \PDN{method-fragment}, are theoretical results in
   themselves.

   \item Having, and developing, a \DN{WSML} will be a benefit to students.
   Both once the library is filled with method fragments, as well as during the
   actual filling and structuring of the library. As parts of
   assignments/projects in the context of courses and/or Bachelor/Master thesis
   projects, students can contribute to the creation and evolution of the
   library.

   \item Practitioners are confronted with a plethora of \PDN{method} and
   \PDN{technique} to model an even wider spectrum of distinct aspects of
   \PDN{work-system}. 

   Having a \DN{WSML} would enable practitioners to better judge which
   \DN{method-fragment} to apply in a specific situation. The \DN{WSML} should
   provide them with a roadmap to find their way around the jungle of methods,
   techniques, tools, etc.
\end{itemize}
The development of the \DN{WSML} extends earlier work on similar
libraries~\cite{2000-Proper-ISEBOK, 2001-Janssen-AML}. Now that our theoretical
perspective on \PDN{work-system} \DN{modeling} has matured, and has become a
core part of the \DaVinci\ series of courses and lecture
notes~\cite{2005-Proper-DaVinciOverview}, we are able to build further on these
initial results.

The notion of \DN{method-fragment} is borrowed from the field of method
engineering~\cite{1996-Rossi-MethodEngineering, 1996-Rolland-MethEng}. This
field of research certainly holds relevance to the \DN{WSML} effort. However,
method engineering aims to develop theories to (situationally) construct full
fledged \DN{system-engineering} \PDN{method}. We are `merely' interested in the
\DN{modeling} aspects of \DN{system-engineering}, with the underlying goal of
obtaining a fundamental insight into \DN{modeling} as such.

In the remainder of this document, we will first discuss
(chapter~\ref{approach}) our approach to the development on the \DN{WSML}. We
then continue (chapter~\ref{types}) with a discussion of a domain model for
the \DN{WSML} which amongst other things will identify the types of
\PDN{method-fragment} that will be stored in the library. This is followed by
the initial classification scheme (chapter~\ref{scheme}) to classify
\PDN{method-fragment}. In subsequent versions of the \DN{WSML} we expect these
latter two to evolve. The version as presented in this document is based on
several sources from literature and some experiences from practice. It will
serve as a starting point to codify/classify pre-existing \DN{modeling}
\PDN{method} and \PDN{technique}. In performing the latter exercise, the
classification scheme is likely to evolve.

      \chapter{Development Approach}
\label{approach}
\Version{07-09-05}

Considering the experimental nature of the \DN{WSML} and the uncertainty about
the precise structures of a classification scheme for the
\PDN{method-fragment}, it is most sensible to take an iterative approach in the
development of the above four elements of the library. Even more, in the
context of several courses in the \DaVinci\ series as well as thesis projects,
students have already done partial classifications of modeling elements. The
WSML project aims to incorporate these results from the past as well.

The basic idea of the \DN{WSML} project is to iteratively create a library of
\PDN{method-fragment}. In creating this library, the following results need to
be created and iterated:
\begin{description}
   \item[Typing of \PDN{method-fragment} --] An overall typing of the kind of
   \PDN{method-fragment} which we will discern.

   The structure of \PDN{method-fragment} will be represented as an
   ORM~\cite{2001-Halpin-ORM} schema together with an explanation of the types
   involved.

   \item[Classification scheme --] A classification scheme which can be used to
   classify the \PDN{method-fragment}.
   
   This classification scheme will be represented as a refinement of the
   previous ORM schema, together with an explanation of the identified
   classification dimensions.

   \item[Classification procedure --] When adding elements to a library, the
   classification should be done in a \emph{reproducable} way.  This requires
   the explicit identification of a procedure that should be followed when
   identifying \PDN{method-fragment} and consequently classifying a given
   \DN{method-fragment}.
   
   \item[Library infrastructure --] A database system based on the
   classification schema, that can be used to actually store the classification
   of the \PDN{method-fragment}. This could be an automated system, but initially,
   this can quite well be a manual (e.g.\ one document per fragment) system.

   \item[Library content --] This is the actual library of classifications of
   modeling elements.
\end{description}
In the development of an initial version the \DN{WSML} we identify four main
iterations:
\begin{enumerate}
   \item[0] As a first step a provisional classification scheme should be
   provided and agreed upon. This document contains the proposed classification
   scheme.

   In addition pre-existing classifications/descriptions of relevant
   \PDN{method-fragment} should be gathered.

   \item[1] The pre-existing classifications should be enriched/transformed to
   match the new classification scheme, and possibly modifications should be
   made to the classification scheme based on the classification schemes used
   in the pre-existing classifications.

   \item[2] Additional \PDN{method-fragment} should be classified and a proper
   library infrastucture should be set up, which is to be filled with the
   classifications.

   \item[3] More \PDN{method-fragment} elements should be classified and the
   classification scheme should be validated against selection criteria as used
   in practical \DN{modeling} situation.
\end{enumerate}
In the further development of the \DN{WSML} we will continue to rely on student
projects. However, we will also seek collaborations with industrial partners,
as well as organizations such as the NGI (Netherlands Society for Informatics),
GIA (Society for Information Architects) and the NAF (Netherlands Architecture
Forum).

      \chapter{WSML Domain Model}
\label{types}
\Version{03-10-05}

In this chapter we discuss the domain model of the \DN{WSML}. The complete
model is depicted in \SmartRef{Concepts}. Below we provide a textual discussion
of the core concepts as identified in the model. Using \SF{sans-serif} font we
will refer to object types in the model depicted in \SmartRef{Concepts}.

\section{Documents}
The library not only comprises facts about \PDN{method-fragment}, but (in
particular in the earlier iterations) comprises documents supporting these
facts. These \SF{Library Source Document}s are \SF{Publications} from the IRIS
publication management system, which may have a \SF{PDF} formatted concent and
which \emph{must} have their \SF{Publication Details} specified as a
\SF{BiBTeX} entry. \SF{Publication}s may refer to other documents.  We
distinguish between two classes of \SF{Library Source Document}s:
\begin{enumerate}
   \item The \SF{Base Document}s are original documents discussing one or more
   method fragments and/or characterisation dimensions, such as Object-Role
   Modeling (ORM)~\cite{2001-Halpin-ORM}, the Zachman
   framework~\cite{1987-Zachman-ISA}, Unified Modeling Language
   (UML)~\cite{1999-Booch-UML}, etc.  
   \item The \SF{Description Document}s are descriptions of \SF{Library
   Fragment}s. As time progresses, the description of a \SF{Library Fragment}
   may be updated. A newer \SF{Description Document} can therefere be preceded
   by an older one.
\end{enumerate}
Initially, the library will be filled mainly with description documents only
(potentially even violating some of the constraints in \SmartRef{Concepts}).
As time goes on, the characterizations of the method fragments will be more and
more explicit, which can then be stored in the library explicitly conform the
structures as described below.

In the library we identify two main classes of \SF{Library Fragment}s. The
first class deals with \SF{Characterization Dimension}s, which are used to
characterize the actual \SF{Method Fragment}s. The latter ones form the other
class of \SF{Library Fragment}s. Below, we discuss these two classes in more
detail, including a further classification of \SF{Method Fragment}s.

\section{Characterizations}
\SF{Method Fragment}s in the library must be characterized by means of 
\SF{Characterization Properties}, which are defined as:
\begin{Definitions}
   A specific value from a \DefinitionName{characterization-dimension}{ls} 
that an artefact may exhibit. 

\end{Definitions}
These characterization properties, which will
be discussed in the next chapter, may refer to such things as the system aspect
(business, information, infrastructure, etc.) which a method fragment is geared
towards, the kind (prescriptive or descriptive) of model it aims to produce,
etc.

Each \SF{Characterization Property} may be associated to a \SF{Method Fragment}
as either being: \SF{intended for} or \SF{suitable for}.  One would expect the
set of properties for which a fragment is intended for to be a subset of
the properties it is suitable for, but that does not necessarily have to
be the case.  In each case, a \SF{Motivation} for the property must be
provided. A \SF{Motivation} may refer to publications (by means of a
\SF{BiBTeXKey}) substantiating this motivation.

A \SF{Characterization Property} really consists of two parts: a
\SF{Characterization Dimension} and a \SF{Characterization Value}. When
characterizing \PDN{method-fragment}, a well defined method should be used in
determining the characterization values. This is described in the
\SF{Characterization Method}. The textual description may refer to other
documents by means of a \SF{BiBTeXKey}. Each \SF{Characterization Dimension}
must be described by some \SF{Dimension Description Document}

Note: What is currently missing from the domain model is the fact that once
a larger set of \SF{Characterization Dimension}s becomes available we can
identify for which types of \SF{Method Fragment}s they are relevant. As a next
step we can then require a characterization of fragments for this set of
dimensions to be mandatory. This will lead to the introduction of a fact-type:
\begin{quote}
   \SF{Characterization dimension (name) is mandatory for Fragment type (name)}
\end{quote}
with constraints:
\begin{quote}
  \SF{EACH Characterization dimension MUST BE mandatory for SOME Fragment type}\\
  \SF{EACH Fragment type IS ONE OF: Viewing framework, Viewpoint, ...}\\
  \SF{EACH Method Fragment of type $X$ is characterized by EACH
  Characterization dimension which is mandatory for type $X$}
\end{quote}

\section{Method fragments}
Each \SF{Method Fragment} must be described by some \SF{Fragment Description
Document}, and must receive some \SF{Characterization Property}s defining what
the intended/suitable focus of the fragment is. We distinguish four main
classes of \SF{Method Fragment}s:
\begin{enumerate}
   \item \SF{Modeling Fragment}s which are the modeling techniques, frameworks,
   viewpoints, types of mappings/relations between models, etc, used in the
   creation of actual models.
   \item \SF{Model}s representing complete models, cases, reference models,
   patterns, etc. 
   \item \SF{View}s comprising sets of models, representing complete cases,
   reference models, patterns, etc.
   \item \SF{Model Relation}s between \SF{Partial Model}s present in the
   library using one of the identified \SF{Model Relation Type}s.
\end{enumerate}
Each of these classes is discussed in more detail in the remainder of this
section.

\subsection{Modeling fragments}
In~\cite{1989-Seligmann-Framework, 1990-Wijers-Modelling} a system development
method is dissected into a \DN{way-of-thinking}, \DN{way-of-modeling},
\DN{way-of-working}, \DN{way-of-controlling} and a \DN{way-of-supporting}.
By combining this terminology with the notion of a viewpoint taken
from~\cite{1999-IEEE-Architecture}, and adding \DN{viewing-framework},
\DN{viewing-cell} and \DN{model-relation-type}, we define the notion of
a \DN{modeling-fragment}, and its constituents, as follows: 
\begin{Definitions}
   A \DefinitionName{method-fragment}{ls} that is concerned with the 
actual \DefinitionName{modeling}{ls} of a \DefinitionName{domain}{ls}. Three classes of 
\DefinitionName{method-fragment}{lp} are identified: \DefinitionName{viewing-framework}{ls}, 
\DefinitionName{way-of-working}{ls}, \DefinitionName{viewpoint}{ls}, \DefinitionName{model-relation-type}{ls}. Each 
\DefinitionName{method-fragment}{ls} must have a unique underlying \DefinitionName{way-of-thinking}{ls}.

   Articulates the assumptions on the kinds of problem domains, solutions,
engineers, analysts, etc. This notion is also referred to as \emph{die
Weltanschauung} \cite{1983-Sol-MethodConsideration, 1985-WoodHarper-Multiview},
\emph{underlying perspective} \cite{1981-Mathiassen-SystemDevelopment} or
\emph{philosophy} \cite{1995-Avison-ISDevMeth}.

   A kind of \DefinitionName{modeling-fragment}{ls}, which identifies a 
set of related \DefinitionName{perspective}{lp} from which to 
\DefinitionName{view}{ls} a \DefinitionName{system}{ls}. A 
\DefinitionName{viewing-framework}{ls} may be composed of yet other \DefinitionName{viewing-framework}{lp}.

   A \DefinitionName{viewing-framework}{ls} that is not decomposed into a 
(covering) set of smaller \DefinitionName{viewing-framework}{lp}.

   Structures (parts of) the way in which a system is engineered.  It defines the
possible tasks, including sub-tasks, and ordering of tasks, to be performed as
part of the development process.  It furthermore provides guidelines and
suggestions (heuristics) on how these tasks should be performed.

   A specification of the conventions for constructing and using \PDN{view}.

This involves:
  a \DN{way-of-thinking}, 
  a \DN{way-of-modeling}, 
  a \DN{way-of-communicating},
  a \DN{way-of-working},
  a \DN{way-of-supporting} and
  a \DN{way-of-using}

   A type of well-defined relations between \DefinitionName{model}{lp}. An 
example would be a mapping algorithm between a conceptual \DefinitionName{model}{ls} and 
a relational database schema, the relation between the \DefinitionName{concept}{lp} of a business 
process \DefinitionName{model}{ls} and the concepts of a supporting work-flow, etc.

   Identifies the \emph{core concepts} of the language that may be used to 
denote, analyze, visualize and/or animate \PDN{system-description}.

   A set of \DefinitionName{modeling}{ls} \DefinitionName{concept}{lp} by 
which \DefinitionName{viewer}{lp} are to observe 
\DefinitionName{domain}{lp}. This usually takes the form of a \DefinitionName{meta-model}{ls}.

   The medium and `notations' used to represent the 
\DefinitionName{concept}{lp} as identified in a \DefinitionName{way-of-conceiving}{ls}.
It describes how the abstract concepts from the \DN{way-of-conceiving}
are communicated to human beings, for example in terms of a textual or a
graphical notation.

Note that it may very well be the case that different \PDN{modeling-technique}
are based on the same \DN{way-of-conceiving}, yet use different notations.

\end{Definitions}
Note: as a synonym for `\DN{way-of-working}' we will also use the term
\DN{approach} and as synonym for `\DN{way-of-modeling}' we will also use
the term \DN{technique}.

\subsection{Models}
The second class of \PDN{method-fragment} deals with \SF{Model}s.
Some may be \SF{partial} models in the sense that
they deal with such things as reference models, reference architectures,
patterns, partial solutions, etc.  All of these latter these models can be
regarded as incomplete in the sense that they all require some form of
\emph{applying} to make them usefull in a specific situation. They cannot be
used directly, but need tuning, extending, or any other form of
\emph{application} to make them fit a situation at hand.

At present, this class of \PDN{method-fragment} is not yet worked out in full
detail, as also shown in \SmartRef{Concepts}. Further research is indeed needed
to ascertain what concepts are involved in \SF{Partial Model}s, their
representation and their application/applicability in specific situations.

\subsection{Views}
A \SF{Viewpoint} comprises one or more \SF{Technique}s. A \SF{View} essentially
is a set of \SF{Model}s describing the same domain in terms of the
\SF{Technique}s associated to the \SF{Viewpoint} used in creating the
\SF{View}. Similarly to \SF{Model}s, \SF{View}s can be partial as well.  If a
\SF{Model} is signified as being a \SF{partial} model, then any \SF{View}
containing this \SF{Model} must also be a \SF{partial} view.

\subsection{Model relations}
The \SF{Model Relation Types} which define relations between models in general
can be used to identify specific relations between \SF{Partial Model}s. This
leads to the notion of a \SF{Model Relation}, which runs \SF{from} a partial
model \SF{to} another one, and is an instance of some \SF{Model Relation Type}.
Similarly to \SF{Partial Model}s, the notion of \SF{Model Relation} will
definitely need further refinements in future versions of the \DN{WSML}.

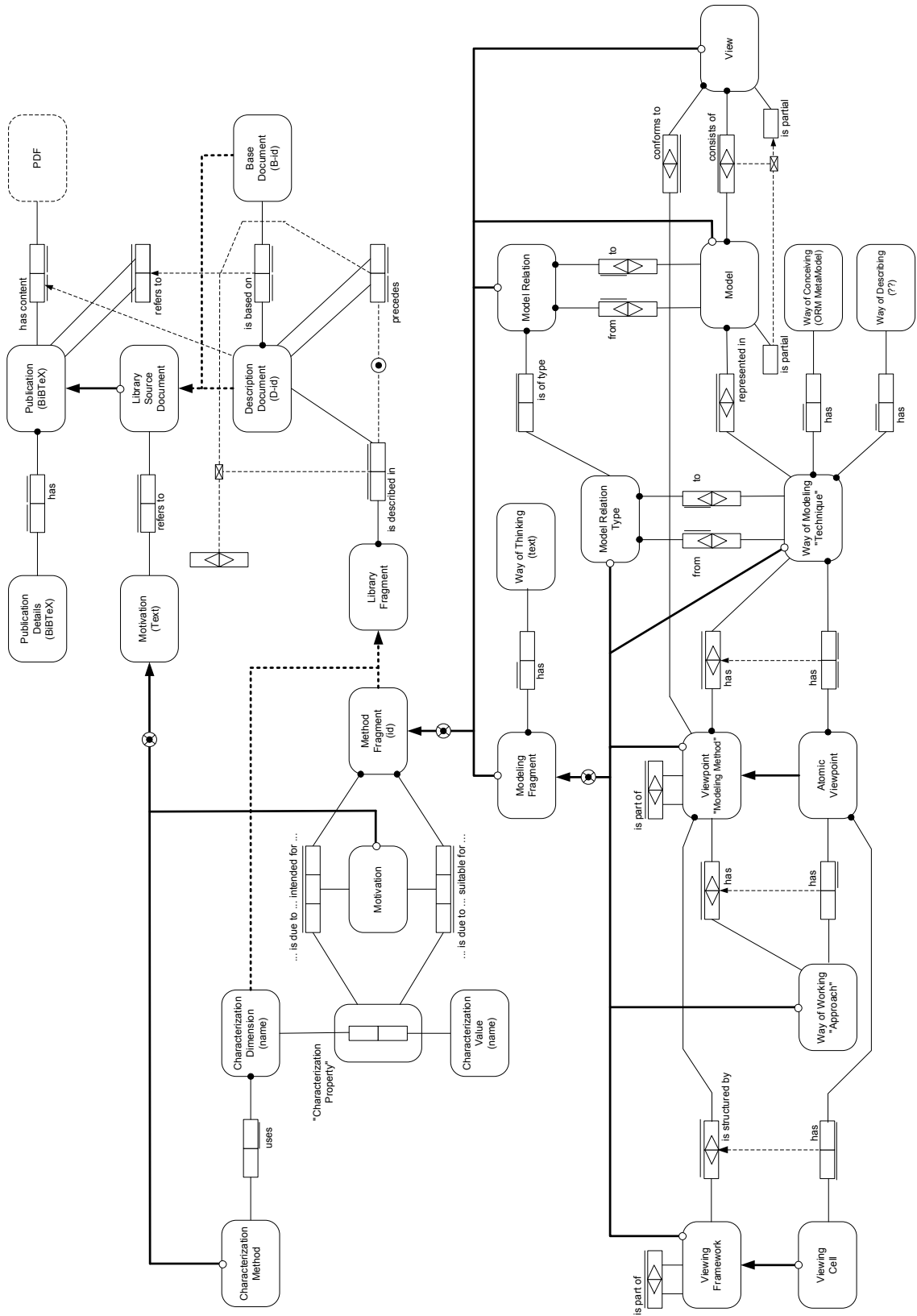
\SizedImage{angle=90,width=\textwidth}{Domain model of the WSML}{Concepts}

      \chapter{Characterization dimensions}
\label{scheme}

In this chapter we provide an initial overview of
\PDN{characterization-dimension} based on existing literature. During the
actual populating of the \DN{WSML}, these dimensions will need to be described
in more detail, in particular the \SF{Characterization Method}. Below, we
provide a first set of classification dimensions. Note that the set of
dimensions used in the \DN{WSML} should be fairly `orthogonal'.

The field of method engineering~\cite{1996-Rossi-MethodEngineering,
1996-Rolland-MethEng} has already done some work on the classification of
(fragments of) methods. This work needs to be taken into account when defining
the \DN{characterization-dimension}. It should be noted, however, that our focus
is on domain modeling methods and not so much on methods for work systems
engineering in general. Even more, our goal of studying modeling methods is to
improve our understanding of the act of modeling as such. In other words, we
are less interested in the construction of methods, but more in understanding
the inner workings of \emph{modeling} methods.

At present we identify four classes of characterization dimensions:
\begin{itemize}
   \item Why the model is produced.
   \item What the model is focussing on.
   \item How the model is produced.
\end{itemize}
Each of these three classes are discussed in detail below. The current set is
based on~\cite{1993-Tapscott-ParadigmShift, 1999-Franckson-ISPL-Deliverables,
1999-Proper-ISPL-LSM, 2001-Janssen-AML, 2002-Heuvel-ArchitectuurPragmatiek,
2003-Greefhorst-Frameworks, 2004-Proper-DaVinci, 2005-Lankhorst-ArchiMate,
2005-Proper-CommunicatingArchitecture}.

\section{Why the model is produced}
\begin{description}
   \item[Modeling purpose --] The purpose for which the models produced by the
   method are suitable for. Based on~\cite{2005-Lankhorst-ArchiMate} we shall
   use as sub-classifiers: 
   \begin{description}
      \item[Designing --] A model which aims to support designers in the design
      process, from initial sketch, via limitations of the design space to the
      detailed design.

      \item[Deciding --] A model which aims to support decision makers in their
      decision making process by offering insight into the design of a domain
      and the possible impact of design decisions.

      \item[Informing --] A model which aims to inform stakeholders about the
      domain being modeled. This is usually done in order to achieve
      understanding, obtain commitment, etc.
   \end{description}
   
   \item[Design chain --] The focus with regard to the design from \emph{why it
   should be} and \emph{what should be} to \emph{how it ought to be} and
   \emph{how it will be}. In other words, the question which the model aims to
   answer about a given work system. Specific methods may produce models that
   are geared towards a specific phase in the design chain.  Based
   on~\cite{2005-Proper-CommunicatingArchitecture} we shall use the following
   sub-classifiers:
   \begin{description}
      \item[Systen purpose --] A model focussing on \emph{why} a specific work
      system is needed.

      \item[System functionality --] A model focussing on \emph{what}
      functionality a work system should provide to its environment.

      \item[System design --] A model representing \emph{how} the functionality
      will be realized.

      \item[System quality --] A model identifying \emph{how well} a work
      system should realize its functionality.

      \item[System costs --] A model defining \emph{what it may cost} for the
      work system to be constructed and deliver the desired functionality to
      its environment.
   \end{description}

   \item[Intended audience --] The audience the model is produced for. Inspired
   by~\cite{2003-Greefhorst-Frameworks} we identify the following
   sub-classifiers:
   \begin{description}
      \item[Actor in future system --] A (human!) actor in a future system.

      \item[Sponsor --] A sponsor of the future system and/or its development.

      \item[Designer --] Designers of the system. 

      \item[Analyst --] Analysts of (parts of) the domain in which the system
      will operate.

      \item[Engineer --] The engineers of the actual system.
   \end{description}
\end{description}

\section{What the model is focussing on}
\begin{description}
   \item[Semantic force --] The semantic force of the model refers to the fact
   wether it is a \emph{prescriptive} or \emph{descriptive} nature, leading to
   sub-classifiers:
   \begin{description}
      \item[Prescriptive --] For models with a purely prescriptive nature.
      These are models which limit the design freedom of future designers who
      are bound to make a design \emph{conforming to} the earlier pre-scriptive
      model.

      \item[Descriptive --] For models with a purely descriptive nature.

      \item[Mixed --] For models with a mix between prescriptive and
      descriptive nature.
   \end{description}

   \item[Nature of the information --] This refers to the nature of the
   content. Based on~\cite{2003-Greefhorst-Frameworks} we identify the
   following sub-classifiers:
   \begin{description}
      \item[Policy --] Policy statements pertaining to the system.

      \item[Principles --] Principles to which the (design of) the system
      should adhere.

      \item[Guidelines --] Guidelines (operationalisations of the principles)
      which should be met by the future system.

      \item[Descriptions --] Descriptions of what the (future) system (will)
      look(s) like.

      \item[Standards -- ] Standards (to be) used in the creation/design of the
      system.
   \end{description}

   \item[Type of information --] This refers to the kind of information that
   may be contained in the model. Inspired by
   e.g.~\cite{1993-Tapscott-ParadigmShift, 2005-Lankhorst-ArchiMate}, typical
   sub-classifiers would be:
   \begin{description}
      \item[Business --] Business models, markets, products, etc.

      \item[Organization --] Work processes, culture, organizational
      structures, skills, etc.

      \item[Information --] Domains of information/knowledge needed for the
      business activities.

      \item[Application --] Automation of work.

      \item[Infrastructure --] Underlying technological infrastructure.
   \end{description}
   Most of the ``architecture frameworks'' provide different sub-classifications
   for these classes of information.

   \item[Systemic scope --] The scope of the domain that is modeled by the
   model. We identify three sub-classifiers:
   \begin{description}
      \item[Use-case --] The scope of a specific use-case of the work system.

      \item[System component --] A distinct component of a work system.

      \item[System --] The entire work system and its direct environment is the
      domain that is to be modeled.

      \item[System of Systems --] The scope is a set of work systems, in other
      words, a system of work systems.
   \end{description}

   \item[Temporal scope --] The temporal scope at which we regard domain.
   Sub-classifiers are:
   \begin{description}
      \item[Operational --] The work system as it is currently (or in the near
      future) operational.

      \item[Tactical --] The work system as it is ideally operational after the
      execution of a development projects.

      \item[Strategical --] The work system as it is ideally operational after
      the execution of a number (a program) of development projects.
   \end{description}

   \item[Implementation abstraction --] The level of abstraction from
   underlying information technology. Based on the distinction from
   MDA~\cite{}, we define the following sub-classifiers:
   \begin{description}
      \item[Computing independent --] A model which is independent of
      decisions regarding computerisation of certain activities.

      \item[Platform independent model --] A model in which choices for
      computerisation of tasks has indeed been made, but does not yet make
      a choice for a specific technological platform.

      \item[Platform specific --] A model which is tied to a specific
      technological platform. A program in a programming language such as C or
      Java would be an ultimate example of a platform specific model.
   \end{description}

   \item[Systemic aggregation --] The level of aggregation that is used in a
   model. Systemic aggregation is usually used as a means to ``hide''
   information about specificities of a work system. It allows modelers (and
   viewers of the models) to focus on higher level issues. Based
   on~\cite{2005-Lankhorst-ArchiMate} we identify the following
   sub-classifiers:
   \begin{description}
      \item[Detailed level --] All (relevant) details are shown.

      \item[Coherence level --] The model focuses on the coherence (between
      different aspects) of the modeled domain.

      \item[Overview level --] The model provides an overview of the key issues
      of the modeled domain.
   \end{description}

   \item[System qualities --] Models may focus on different \PDN{quality-property}
   of systems. In \cite{1991-ISO-Quality, 1996-ISO-KSP} the following key
   \PDN{quality-attribute} have been identified:
   \begin{Definitions}
     Is the relationship between the level of performance of the \DN{system} (or a
\DN{sub-system}) and the amount of resources used, under stated conditions.
Efficiency may be related to time behaviour (response time, processing time,
throughput rates) and to resource behaviour (amount of resources used and
duration of such use).

     Bears on the existence of a set of functions and their specified properties.
The functions are those that satisfy stated or implied needs.

Sub-characteristics of functionality are: suitability, accuracy,
interoperability, compliance, security and traceability. 

     Is the capability of the \DN{system} (or a \DN{sub-system}) to maintain its
level of performance under stated conditions for a stated period of time.  The
mean time to failure metric is often used to assess reliability of systems.
The reliability can be determined through defining the level of protection
against failures and the necessary measures for recovery from failures.

Sub-characteristics of reliability are: maturity, fault tolerance,
recoverability, availability and degradability.

     Bears on the effort needed to make specified modifications to the \DN{system}
(or a \DN{sub-system}).  Modification may include corrections, improvements or
adaptation to changes in  the environment, the requirements and the (higher
levels of the) design.

Sub-characteristics of maintainability are: analysability, changeability,
stability, testability, manageability, reusability.

     Is the ability of the \DN{system} (or one of its \PDN{component}) to be
transferred from one environment to another.

Sub-characteristics of portability are: adaptability, installability,
conformance, replaceability.

     Bears on the effort needed for the use of the \DN{system} (or one of its
\PDN{sub-system}) by the actors in the \DN{environment} of the system.

Sub-characteristics of usability are: understandability, learnability, operability, 
explicitness, customisability, attractivity, clarity, helpfulness and 
user-friendliness.

   \end{Definitions}

   \item[System realization --] The level of realization of the services
   provided by a system to its environment. The underlying way of thinking is
   that a work system is a \emph{supporting system}, which provides
   functionality to a \emph{using system}, while making use of
   \emph{infrastructure system(s)}. Sub-classifiers are therefore:
   \begin{description}
      \item[Using system --] The way the environment will use the work system.

      \item[Supporting system --] The way the work system will provide
      functionality/services to the environment.

      \item[Infrastructure system --] The infrastructural facilities that are
      used by the work system in order to provide the functionality/services to
      the environment.
   \end{description}
  
   \item[Actor kinds --] What kind of actors \emph{in the system} does the
   modeling focus on?  Sub-classifiers are:
   \begin{description}
      \item[Heterogenous --] Heterogenous (typically composed) actors.

      \item[Human --] The role of human beings in a work system.

      \item[Computerised intelligence --] The role of software
      agents/components that aim to show intelligent behavior.

      \item[Computerised --] The role of ``traditional'' software components.
   \end{description}

   \item[Explicitness of represented knowledge --] Models produced during system
   development are essentially forms of \emph{explicit knowledge} pertaining to
   an existing/future system; its design, the development process by which it
   was/is to be created, the underlying considerations, etc.  
   
   Based on~\cite{1999-Franckson-ISPL-Deliverables, 1999-Proper-ISPL-LSM}, the
   following dimensions of explicitness for representations of system
   development knowledge (pertaining to both target domain and project domain
   knowledge) can be
   identified: 
   \begin{description}
     \item[Level of formality --] The degree of formality indicates the type of
     language used to represent the knowledge. Such a language could be formal,
     in other words a language with an underlying well-defined semantics in
     some mathematical domain, or it could be informal --not mathematically
     underpinned, typically natural language, graphical illustrations,
     animations, etc.
   
     \item[Level of quantifiability --] Different aspects of the designed
     artefact, be it (part of) the target or the project domain, may be
     quantified. Quantification may be expressed in terms of volume, capacity,
     workload, effort, resource, usage, time, duration, frequency, etc.
     
     \item[Level of executability --] The represented knowledge may, where it
     concerns artefacts with operational behaviour, be explicit enough so as to
     allow for execution. This execution may take the form of a simulation, a
     prototype, generated animations, or even fully operational behaviour based on
     executable specifications.
   
     \item[Level of comprehensibility --] The knowledge representation may not
     be comprehensible to the indented audience. Tuning the required level of
     comprehensibility of the representation, in particular the representation
     language used, is crucial for effective communication. The representation
     language may offer special constructs to increase comprehension, such as
     stepwise refinements, grouping/clustering of topically related
     items/statements, etc.
   
     \item[Level of completeness --] The knowledge representation may be
     complete, incomplete, or overcomplete with regards to the knowledge topic
     (see previous subsection) it intends to cover.
   \end{description}
\end{description}
Note: models may not only focus on one of the areas identified in the
sub-classifiers, they are also likely to identify links between the different
areas, leading to the vertical/horizontal integration of models with a
homogeneous focus. This will need some refinement of the classification scheme.

\section{How the models are produced}
Based on~\cite{1999-Franckson-ISPL-Deliverables, 1999-Proper-ISPL-LSM}, we
currently identify:
\begin{description}
   \item[Cognitive approach --] The cognitive approach used during the
   analysis phase represents the way in which information is processed to make
   design decisions in the migration project. Two options are distinguished:
   \begin{description}
      \item[Analytical approach --] When information is processed analytically,
      the available information is simplified through abstraction in order to
      reach a deeper and more invariant understanding.  An analytical approach
      is used to handle the complexity of the information.  In an analytical
      approach the information system is mainly described by use of some degree
      of formality.

      \item[Experimental approach --] When using an experimental approach the
      project actors learn from doing experiments.  The purpose is to reduce
      uncertainties by generating more information.  Experiments can, for
      example, be based on prototypes, mock-ups, benchmark test of migrated
      components or other kinds of techniques which make the results of
      migration scenarios visible.
   \end{description}
   The two cognitive approaches may be combined.  An analytical approach to
   reduce complexity may introduce new sources of uncertainty requiring
   experimental actions. Conversely, an experimental approach to reduce
   uncertainty may introduce new sources of complexity requiring analytical
   actions.  
   
   \item[Social approach --] The social approach is the way in which project
   actors work together with the business actors during the analysis phase.
   Two options are distinguished:
   \begin{description}
      \item[Expert-driven --] In an expert-driven approach, project actors (the
      experts) will produce descriptions on the basis of their own expertise,
      and interviews and observations of business actors.  The descriptions can
      then be delivered to the business actors for remarks or approvals.

      \item[Participatory --] In a participatory approach, the project actors
      produce the descriptions in close co-operation with some or all the
      business actors, e.g.\ in workshops with presentations, discussions and
      design decisions.  A participatory approach may allow the acquisition of
      knowledge, the refinement of requirements and the facilitation of
      organisational change.
   \end{description}
\end{description}

      \Biblio
      \Dictionary
      \AuthorIndex
      \SubjectIndex
   \makeatletter
\ifodd\c@page
\else
   \mbox{~~~}
   \newpage
\fi
\makeatother

\pagestyle{empty}
\stepcounter{page}

\begin{titlepage}
   \setcounter{page}{20001}

   \vspace{3cm}

   {\Large The \DaVinci\ Lecture Notes Series:}

   \begin{quote}
      The \DaVinci\ series of lecture notes is concerned with \emph{The Art \&
      Craft of Information Systems Engineering}. On the one hand, this series
      of lecture notes takes a fundamental view (\emph{craft}) on the field
      information systems engineering. At the same time, it does so with an
      open eye to practical experiences (the \emph{art}) gained from
      information system engineering in industry.
   \end{quote}

   \vspace{1cm}

   {\Large Main contributors:}
   \def\Person#1#2{
     \begin{tabular}[b]{c}
        \ImageBox{2.5cm}{#1}\\
	{\small #2}
     \end{tabular}
   }
   \begin{center}
      \begin{tabular}{cc}
         \Person{patrick}{P. (Patrick) van Bommel} &
         \Person{stijn}{S.J.B.A. (Stijn) Hoppenbrouwers} 
	 \\~~\\
	 \Person{ger}{G.F.M. (Ger) Paulussen}
	 \\~~\\
	 \Person{erik}{H.A. (Erik) Proper} &
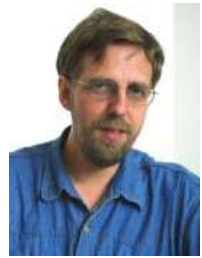
         \Person{theo}{Th.P. (Theo) van der Weide} 
      \end{tabular}
   \end{center}
\end{titlepage}

\end{document}